# High-efficient and polarization independent edge coupler for thin-film lithium niobite waveguide devices

**CHANGRAN HU,[1] AN PAN,[1] TINGAN LI,[1] XUANHAO WANG,[1] YUHENG LIU,[1] SHIQI TAO,[1] CHENG ZENG, [1,2] AND JINSONG XIA[1,3]**

[1]*Wuhan National Laboratory for Optoelectronics, Huazhong University of Science and Technology, Wuhan 430074, China*
[2]*Corresponding author: zengchengwuli@hust.edu.cn*
[3]*Corresponding author: jsxia@hust.edu.cn*

**Abstract:** Lithium niobate (LN) devices have been widely used in optical communication and nonlinear optics due to its attractive optical properties. The emergence of thin-film lithium niobate on insulator (LNOI) improves performances of LN-based devices greatly. However, a high-efficient fiber-chip optical coupler is still necessary for the LNOI-based devices for practical applications. In this paper, we demonstrate a highly efficient and polarization-independent edge coupler based on LNOI. The coupler, fabricated by standard semiconductor process, shows a low fiber-chip coupling loss of 0.54 dB/0.59 dB per facet at 1550 nm for TE/TM light respectively, when coupled with ultra-high numerical aperture fiber (UHNAF) of which mode field diameter is about 3.2 μm. The coupling loss is lower than 1dB/facet for both TE and TM light at wavelengths longer than 1527nm. A relatively large tolerance for optical misalignment is also proved. The coupler shows a promising stability in high optical power and temperature variation.

## 1. Introduction

Over the last two decades, significant progress in the photonic integrated circuits (PICs) based on Si have been made [1,2]. They are crucial for the performance of various optical transmission systems. Recently, the LNOI attracts more intention from both industry and research institutions and is considered as a promising candidate for new-generation PICs platform due to its potential for ultra-high-speed application. Recently, various devices based on LNOI have been reported and shown outstanding performance, such as low-loss optical waveguide, high-Q ring resonator [3], tunable filter [4], high speed EO modulator [5], optical frequency comb [6], second harmonic generator [7], wavelength convertor [8], spectrometer [9], etc. However, a highly efficient, polarization independent and wide-band fiber-chip optical coupler is still absent, which is critical essential for these devices to be used in practical applications.

Unfortunately, the huge difference between mode-field-diameter (MFD) of standard-single-mode-fiber (SSMF, ~10μm) and integrated waveguide on LNOI(<1μm) leads to a great optical loss with directly-coupling configuration [10]. To realize fiber-chip coupling, various optical couplers, such as grating coupler [11], prism coupler [12], horizontal edge coupler [13–23] can be used. The grating coupler is limited to its limited bandwidth and polarization-dependence. As to prism coupler, the requirement of stable mechanical pressure to hold the prism in right place makes it less useful in practical applications [24]. Evanescent coupling is another intuitive scheme utilizing the energy transfer between two closed optical waveguides [25-27], but usually at the cost of the special tapered fibers and uncovered parts of devices, as well as the difficulty in optical packaging. Researchers have also proposed novel optical couplers made by polymer utilizing 3D direct laser writing (DLW), which is based on 3D-printed freeform micro-lenses that are connected to on-chip waveguides by photonic wire bonds (PWB) [28,29]. But the mechanical and thermal stability for the suspended polymer structures are still needed to be proved. Comparing with these coupling schemes mentioned above, horizontal edge couplers [20] better meet the requirements of practical applications due to its advantages of high coupling efficiency, wide band, polarization independence, as well as stability and ease of packaging.

Recently, various kinds of edge couplers based on LNOI have been reported. For example, Chemical-mechanical polishing (CMP) to is utilized to fabricate waveguide covered with tantalum oxide [13], but it is hardly compatible with general ridge waveguide. Mode matching between ridge waveguide and tapered and lensed fiber (TLF) through a single ridge taper [14] is also demonstrated, but the coupling loss is still relatively high [5,8,9]. To solve these problems, bilayer taper couplers haves been proposed to improve the performance [15]. But one is difficult to define the shape and effective index of the mode profile at the facet of the coupler or to prevent the energy leakage into the substrate beneath the buried oxide (BOX) without the cladding waveguide (CLDWG). Recently, researchers reported an edge coupler made up of polymer showing a coupling loss of 1.5dB per facet [30]. But the existence of polymer in the coupler structure can bring mechanical and thermal instability, thereby limiting its application scenarios.

In this paper, we demonstrate a highly efficient, wide-band and polarization independent edge coupler based on LNOI fabricated standard semiconductor process. It shows relatively large tolerance of fiber-chip misalignment and input power stability. The measured UHNAF-to-chip optical coupling loss of 0.54 dB/facet (0.59 dB/facet) for TE(TM) at 1550 nm is achieved. The coupling loss is lower than 1dB/facet for both TE and TM light at wavelengths longer than 1527nm. For TE light, the coupler's loss increases by 0.41 dB when the input power is increased to 30 dBm. An increase of 0.6 dB per facet in loss is induced by a 1-μm misalignment between the fiber end and coupler along X/Z direction. Besides, the thermal stability of the coupling by fitting and adhere the UHNAF using UV curing adhesive is also impressive. The max degradation of coupling efficient for TE light is less than 0.01 dB when cycling the temperature from -35 ℃ to 75 ℃ in 2.5 hours.

## 2. Design and simulation

The schematic structure of the coupler is shown in Fig. 1(a) and (b). The cross-section (CS) view and corresponding mode field distribution of at different points of the coupler are shown in Fig. 1(c). It consists of two parts, a bilayer LN taper to convert the mode adiabatically, and a silicon oxynitride (SiON) CLDWG. The refractive index of SiON is slightly higher than silicon dioxide to define the shape and effective index of the mode profile at the facet of the coupler, while preventing the energy leakage into the substrate beneath the buried oxide (BOX). The width and height of the ridge waveguide (RWG) are designed to be 0.9 μm and 0.26 μm on a 0.24-μm-thick slab to operate in single-mode-condition. Bilayer LN taper is divided into three sections. Firstly, LN slab region narrows linearly while keeping the waveguide width unchanged. Secondly, the width of lower LN slab and width of the ridge reduce simultaneously to form the upper inversed taper. Finally, the lower slab width reduces to form the lower inversed taper. In fact, to obtained high coupling efficiency the tip width of lower inversed taper should be small enough, which is limited by the fabrication process. A cladding SiON waveguide covers the whole taper area. The chip is protected by a $SiO_2$ layer deposited by PECVD (Cladding $SiO_2$ is omitted in Fig.1).

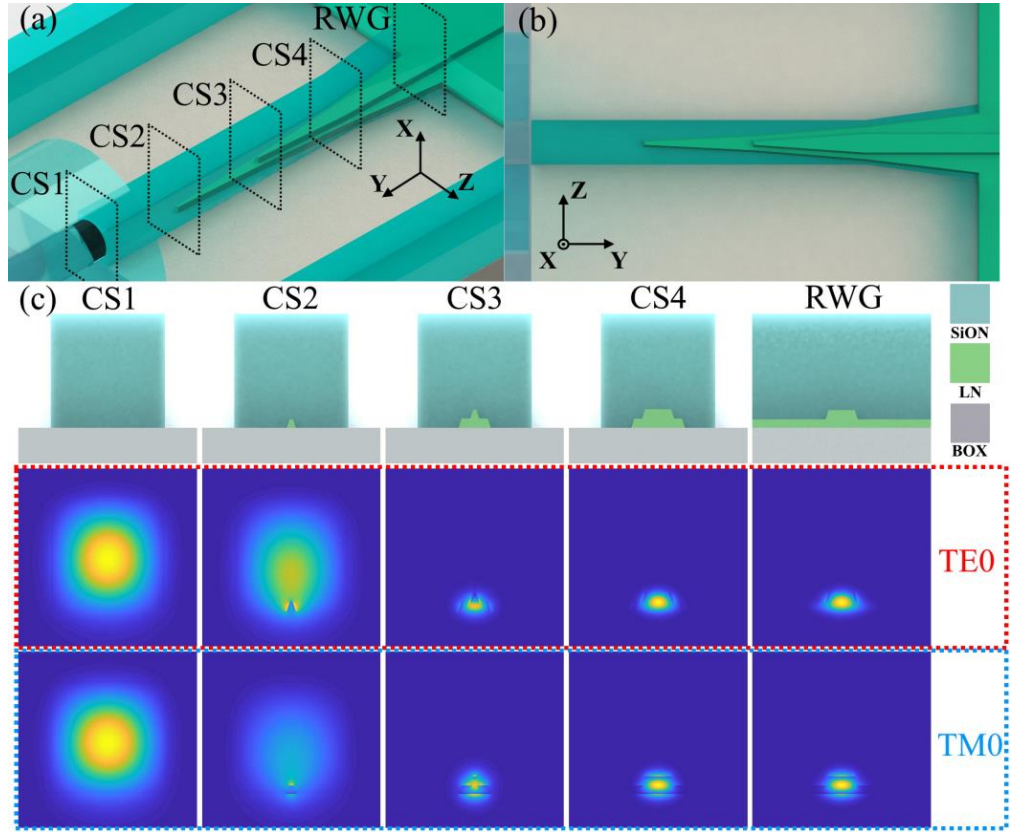

Fig.1. (a) Schematic structure of the edge coupler. (b) Top view of the edge coupler. (c) Cross section views of the coupler and corresponding Ez and Ex distribution of fundamental TE and TM mode, respectively.

The coupler is designed to couple with an UHNAF with mode field diameter (MFD) of 3.2 μm. We simulated the fundamental mode of CLDWG to optimize the parameters, including the size of the CLDWG and material index. A coupling loss of 0.07 dB (0.06 dB) per facet is obtained for TE(TM) light in simulations after optimization. Cross-sectional distributions of the electric field of fundamental TE and TM mode in different points of the coupler are indicated in Fig. 1(c).

As shown in Fig. 1(c), CS1 shows the TE0/TM0 mode profiles supported by the SiON CLDWG, which are similar to the mode supported by the UHNAF with a 3.2-μm MFD. Therefore, the light can be coupled efficiently into the CLDWG from the UHNAF. Then, the modes are gradually "absorbed" into the lower LN taper as shown in CS2 and CS3. After that, the mode continues to transform adiabatically into the fundamental mode supported by the normal LN ridge waveguide through the upper inversed LN taper.

### 3. Fabrication and measurement

We fabricated the coupler on a commercially available X-cut LNOI substrate (NANOLN). The top LN thin film is 500-nm thick, and the buried oxide is 4.7-µm thick. Firstly, The LN ridge waveguide was defined using electron beam lithography (EBL). Then the pattern was transferred into LN thin film through a 260-nm-deep try etch using inductively coupled plasma-reactive ion etching (ICP-RIE). Secondly, the lower taper was patterned in the same way. Then, the CLDWG was defined by depositing 3.2-µm SiON using plasma enhanced chemical vapor deposition (PECVD) followed by dry-etching after lithography. Finally, end face polishing was conducted to form the facet of the coupler after SiO$_2$ covered onto the whole chip. Optical and SEM images of the CLDWG and LN taper are shown in Fig. 2.

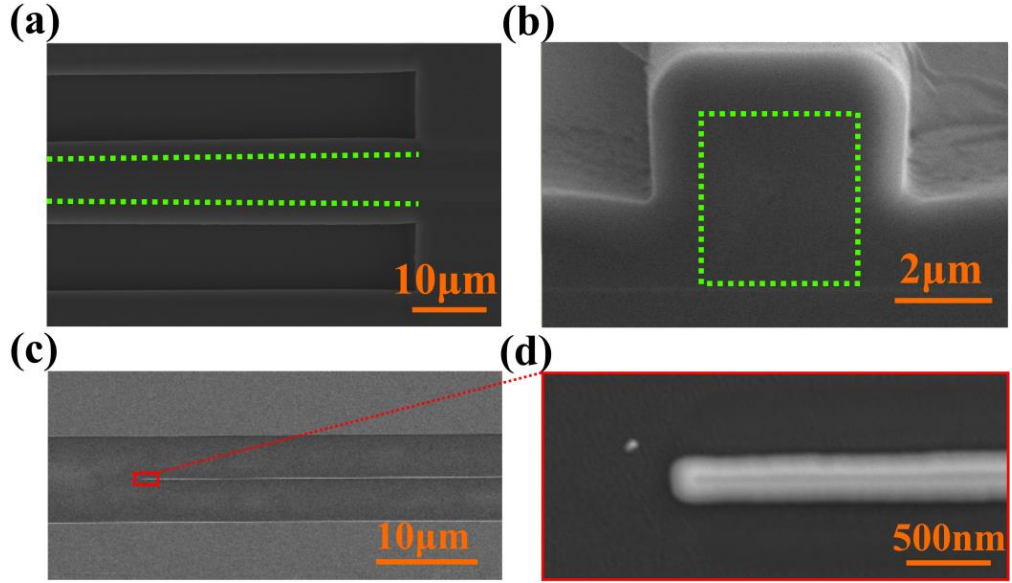

Fig.2. (a) Scanning electron microscope image of the spot size convertor. The CLDWG is marked by the green dotted lines. (b) Scanning electron microscope image of the polished facet of the coupler. The CLDWG is marked by the green dotted lines. (c) Scanning electron microscope image of the lower inversed LN taper. (d) Scanning electron microscope image of the tip of lower inversed LN taper.

A tunable laser (Santec TLS-510) is used as the light source in the measurement of optical coupling test. The light from the UHNAF with a flat head was coupled into the chip after amplified by an erbium-doped fiber amplifier (EDFA) and polarized by a polarization controller (PC). Index matching oil with an index of 1.46 was used to match the index between the UHNAF and the edge coupler to reduce reflection. The light was coupled out from the chip using the same configuration and finally detected by an optical power meter (YOKAGAWA AQ2211). The transmission spectrum of the coupler is obtained by continuous scanning the wavelength from 1500 nm to 1600 nm. The configuration of the coupling test system is shown in Fig. 3.

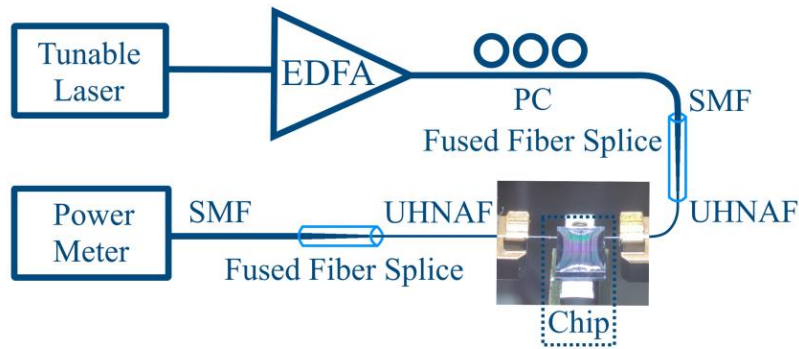

Fig.3. The configuration of the coupling test system.

Transmission spectrum is shown in Fig. 4(a), coupling loss was evaluated to be 0.54 dB/facet and 0.59 dB/facet at 1550 nm for TE and TM light. And the polarization extinction ratio is low than 0.39dB at 1550nm. The coupling loss is lower than 1dB/facet for both TE and TM light at wavelengths longer than 1527nm. The loss decreases when the wavelength increases since it is equivalent to shrinking the size of the tapers' tips. The coupler is able to achieve a coupling loss lower than 1dB/facet in 1530 to 1600 nm. And it is expected that the coupler has a coupling loss lower than 1dB/facet in the whole C+L-band (1530 to 1625 nm) according to the curve trend of the transmission spectrum. An accelerating upward trend of loss is observed when the wavelength decreases. One reason is that the tip width is not small enough for the shorter wavelengths. Another possible reasonable reason is the optical absorption of plasma deposited SiON at around 1510 nm [31-33]. The widening of coupler's operating wavelength might be achieved by changing the material of CLDWG, inserting a SiO2 spacer layer between LN taper and SiON waveguide [33], and reducing the tip width further.

Couplers with various tip widths of the lower LN inversed taper were fabricated and tested. the simulated and measured coupling loss at 1550 nm versus different tip widths taper were shown in Fig. 4(b). As expected, the loss is rising when the tips get wider, while the loss is still as low as 0.72 dB when the tip width reaches 400 nm. This result shows the large tolerance of fabrication error and relatively low requirement of resolution for our proposed coupler. The difference of the coupling loss between experiment and simulation might come from the imperfections induced in fabrication process.

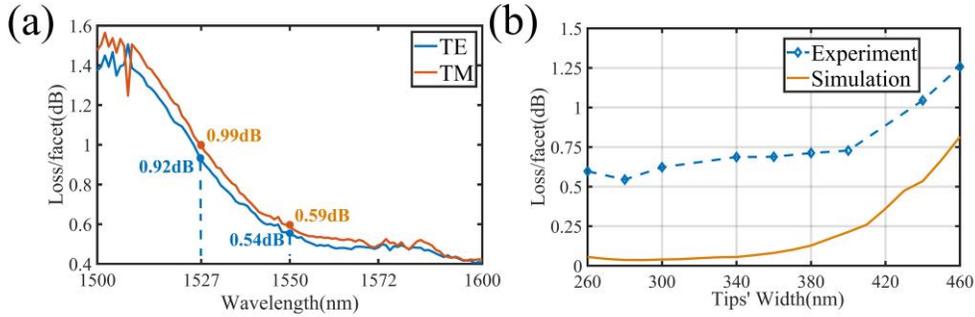

Fig.4. (a) Transmission spectrum of the fabricated coupler. (b) Coupling loss versus different tip widths of the lower LN inversed taper (TE mode).

The tolerance of the alignment between the coupler and the fiber is measured through shifting the fiber from the optimum coupling position along X and Z direction, as shown in Fig. 5(a). For TE light, 1μm of misalignment between fiber and coupler along X/Z direction leads to a 0.6 dB additional loss per facet. Our coupler shows a relatively large tolerance of misalignment, which will reduce the cost of optical packaging and increase the reliability of LNOI devices in practical applications. The tolerance of our coupler induced by its relatively large mode spot size.

High optical power is necessary in some situations such as nonlinear optics application. However, high optical power leads to degradation of coupling efficiency due to the change of refractive index induced by photo damage and photothermal effect. Fig. 5(b) shows the coupling loss at different input powers. The high-power light is provided by an EDFA. The loss increases by about 0.4 dB when the input power is increased to 30 dBm (1 W), while it remains lower than 1dB/facet making it promising for high power applications.

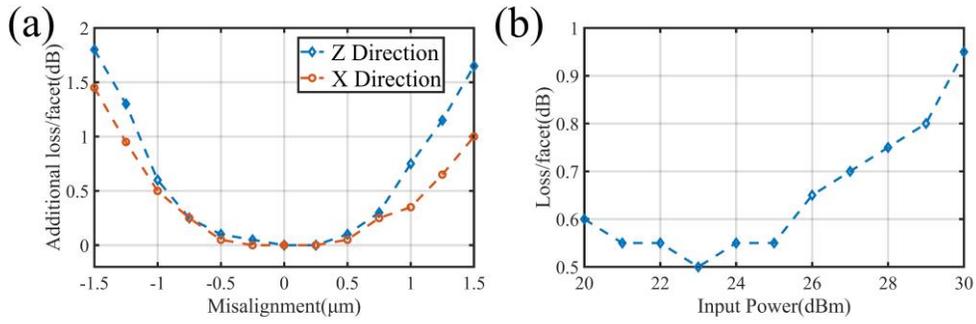

Fig.5. (a) Dependence of coupling loss on fiber-chip misalignment (TE mode). (b)Dependence of coupling loss on optical input power (TE mode).

After fixing the fiber array for UHNAF with the chip by UV curing adhesive, we study the dependence of the coupling loss on the environment temperature in the range of -45 ℃ to 75 ℃. The junction between fiber array for UHNAF and the chip is shown in Fig. 6(a). And the result of thermal loop test is shown in Fig. 6(b). It shows a very low sensitivity to environment temperature. The max degradation of coupling efficient for TE light is less than 0.01 dB. The impressive result of thermal loop indicates the potential of our coupler to for industrial applications.

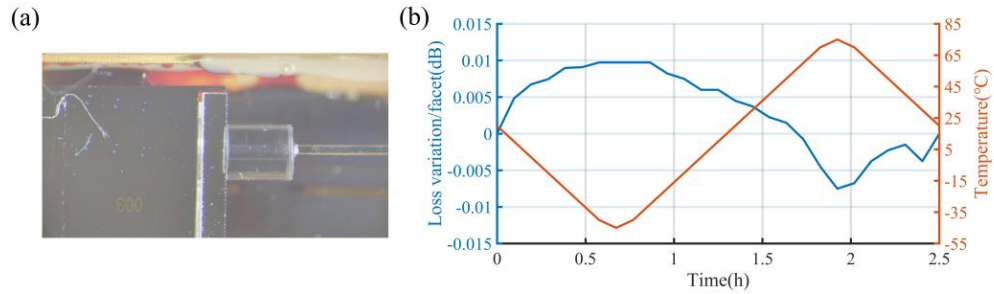

Fig.6. (a) The junction between fiber array for UHNAF and the chip. (b) The dependence of coupling loss on environment temperature (TE mode).

## 4. Conclusion

In this paper, we demonstrated an edge coupler for LNOI devices through combining a bilayer inversed taper and CLDWG. The proposed coupler shows high efficiency, polarization independence and high stability. The coupler is fabricated by standard semiconductor process, making it possible for massive production at low cost. A low coupling loss of 0.54 dB/0.59 dB per facet for TE/TM light is achieved. The coupling loss is lower than 1dB/facet for both TE and TM light at wavelengths longer than 1527nm. A relatively large tolerance for optical misalignment is also proved. The coupler shows a promising stability in high optical power and temperature variation. The demonstrated edge coupler will push LNOI devices closer to practical applications.

**Funding.** National Key Research and Development Program of China (2019YFB2203501), and the National Natural Science Foundation of China under Grant No. 61835008, 61905079, 61905084.

**Acknowledgments.** This work is supported by National Key Research and Development Program of China (2019YFB2203501), and the National Natural Science Foundation of China under Grant No. 61835008, 61905079, 61905084. We thank the Center of Micro-Fabrication and Characterization (CMFC) of WNLO and the Center for Nanoscale Characterization &

Devices (CNCD), WNLO of HUST for the facility support. And we genuinely appreciate the technical support of Doctor Wenbao Sun.